\documentclass[12pt]{iopart}

\usepackage{color}

\usepackage{amssymb}
\usepackage{graphicx}
\usepackage{tikz}

\newcommand{\ret}{x}
\newcommand{\Ret}{X}
\newcommand{\Prop}{G}
\newcommand{\prop}{g}
\newcommand{\eps}{\varepsilon}
\newcommand{\Eps}{\mathcal{E}}
\newcommand{\Cret}{\Sigma}
\newcommand{\cret}{\sigma}
\newcommand{\Csgn}{C}
\newcommand{\csgn}{c}
\newcommand{\resp}{R}
\newcommand{\diag}{\mathrm{diag}}
\newcommand{\off}{\mathrm{off}}
\newcommand{\avg}{\mathbb{E}}

\usepackage[numbers,sort&compress]{natbib}
\usepackage{stmaryrd}
\usepackage{lscape}
\usepackage{tikz}

\begin{document}

\title[Dissecting cross-impact on stock markets:\\  An empirical analysis]{Dissecting cross-impact on stock markets:  An empirical analysis}

\author{M~Benzaquen,  I~Mastromatteo, Z~Eisler and J-P Bouchaud}

\address{Capital Fund Management, 23 rue de l'Universit\'e, 75007 Paris}
\ead{zoltan.eisler@cfm.fr}
\vspace{10pt}
\begin{indented}
\item[]\today
\end{indented}

\begin{abstract}
The vast majority of market impact studies assess each product individually, and the interactions between the different
order flows are disregarded. This strong approximation may lead to an underestimation of trading costs and possible contagion effects. Transactions 
in fact mediate a significant part of the correlation between different instruments.
In turn, liquidity shares the sectorial structure of market
correlations, which can be encoded as a set of
eigenvalues and eigenvectors. We introduce a multivariate linear propagator model that successfully describes such a structure,
and accounts for a significant fraction of the covariance of stock returns. We dissect the
various dynamical mechanisms that contribute to the joint dynamics of assets. We also define two simplified models with substantially
less parameters in order to reduce overfitting, and show that they have superior out-of-sample performance.
\end{abstract}

% Uncomment for PACS numbers
%\pacs{00.00, 20.00, 42.10}
%
% Uncomment for keywords
%\vspace{2pc}
%\noindent{\it Keywords}: Market impact, cross-impact, US stocks.
%
% Uncomment for Submitted to journal title message
%\submitto{\JPA}
%
% Uncomment if a separate title page is required
%\maketitle
% 
% For two-column output uncomment the next line and choose [10pt] rather than [12pt] in the \documentclass declaration
%\ioptwocol

\section{Introduction}
Price impact in financial markets -- the effect of transactions on the
observed market price -- is of both scientific and practical
relevance~\cite{bouchaud2008markets}. A long series of studies has
concentrated on its various aspects in the past
decades~\cite{hasbrouck1988trades,hasbrouck1991measuring,jones1994transactions,dufour2000time,bouchaud2004fluctuations,lillo2004long,eisler2012price,bacry2014hawkes}. The
metrics used in this body of work are usually
calculated individually on each product, and possibly averaged across them
afterwards. The interactions between their
order flows are typically disregarded. This is a very strong approximation, given 
that a financial instrument is rarely traded on its own. Most investors
construct diversified portfolios by buying and selling tens or even hundreds
of assets at the same time. Some of these might be similar, or even almost
equivalent to each other (companies in the same industrial sector,
dual-listed shares, etc.). In these cases it is immediately clear
that to treat each of them separately is not justified, and
often an underestimation of impact costs.
Intuition tells us that in two related products the order flow of one of them
may reveal information, or communicate excess supply/demand regarding
the other. How important are such effects, both
qualitatively and quantitatively?

The ``self-impact'' of a product's order flow
on its own price, as studied in the literature, is an important
component of price dynamics. In comparison, is ``cross-impact'' a
detectable effect? If it is, is it strong enough to significantly
contribute to cross-correlations between stocks? This question was already
raised in the seminal work of Hasbrouck and Seppi \cite{hasbrouck2001common}.
It is particularly interesting, because in spite of the importance of
cross-correlations in  risk management, their microstructural origin is
not clear. Many partial, competing explanations exist, for a review
of recent economics literature on the subject see in Ref. \cite{boulatov2013informed}.
When choosing their quotes, liquidity providers use
correlation models calibrated from real data. It would thus be a
circular argument to fully ascribe such correlations emerge to market makers' quote adjustments. 
A dynamical explanation is more plausible. When two stocks get out of
line relative to one another, liquidity takers may also act on such a mispricing. 
As they consume liquidity, market makers adjust their pricing to avoid building up a
large inventory: this is price impact. As the relative price reaches
a (temporary) market consensus order flows become balanced. Several structural, equilibrium theories exist
with such dynamics, but the underlying models often have many
parameters which cannot be directly fitted to data. Only the
qualitative predictions can be observed, which are nevertheless
very important for practical purposes \cite{pasquariello2013strategic}.
 
In this manuscript we argue in favor of such a dynamical
picture, where transactions mediate a significant part of the
interaction between different instruments, and price impact is an
integral part of price formation. We will demonstrate quantitatively
that correlations and liquidity are intertwined. 
Refs.~\cite{wang2016average,wang2016cross} revisit the evidence for
cross-impact by analyzing the cross-correlation structure of price
changes and order flows. Our study complements such a perspective by
focusing on the underlying
interactions rather than on correlations. Based on a variant of
the well-known propagator technique \cite{bouchaud2004fluctuations},
calibrated on anonymous data, we
will show that liquidity displays a sectorial structure related to the
one of market correlations, that we will be able to describe through
decomposition in eigenvalues and eigenvectors.
This is in the spirit of the principal component analysis
approach advocated in Ref.~\cite{hasbrouck2001common}, and the
analysis of Ref.~\cite{pasquariello2013strategic} from an
econometric point of view.

For the sake of simplicity we will use here the language of
stocks, and we will in fact limit our datasets to these.
However, the techniques introduced below can be applied to many other
markets.  Moreover, note that we focus here on the impact of the
aggregated order flow, rather than the one of a \emph{meta-order} (a
sequence of trades in the same direction submitted by the same actor).
Even though the propagator formalism that we employ is known to 
predict inaccurately the impact of a meta-order, it still provides
qualitatively reliable estimates of market impact (see Section~5
of Ref.~\cite{mastromatteo2014agent}). Thus, we believe that the
cross-interaction network that we find should generalize to the
meta-order case as well, at least to a good approximation.

The paper is structured as follows. Section
\ref{sec:dataset-description} introduces basic notations and our
dataset. Section \ref{sec:market-impact-price} defines a few
fundamental quantities related to returns and price impact, and
summarizes that main stylized facts that we observe. Section
\ref{sec:simple-model} provides a non-parametric
multivariate propagator model, which
is then fitted to the data. Section \ref{sec:dimensional_reduction}
analyzes simpler, lower-dimensional models that can more efficiently
capture the structure of cross-impact; and
compares their in-sample and out-of-sample performance. Finally,
Section \ref{sec:conclusions} concludes.

%%%%%%%%%%%%%%%%%%%%%%%%%%%%%%%%%%%%%%%%

\section{Data and notations}
\label{sec:dataset-description}

We conduct our empirical analysis on a pool of $N=275$ US stocks
as representative as possible in terms of liquidity, market capitalisation
and tick size. The large number of assets and their diversity ensures strong
statistical significance, and allows us to investigate the scaling of
our results when the number of products becomes large. The data consists of five-minute binned trades and
quotes information from January 2012 to December 2012, extracted from
the primary market of each stock (NASDAQ or NYSE). Furthermore, we only
focus on the continuous trading session, removing systematically
the first hour after the open and the last 30 minutes before the
close. In this way we avoid artifacts arising from the particularities
of trading activity in these periods. Out-of-sample tests will be carried out
on an equivalent dataset from 2013. \smallskip

For each five-minute window whose end point is $t$ and for each asset $i$, we compute the
{log-return} $ \ret^{{i}}_t = \Ret^{{i}}_t - \Ret^{{i}}_{t-1}$, where  $\Ret^{{i}}_t =
\log p^{{i}}_t $ and where $p^{{i}}_t$ denotes the price of stock $i$ at time
$t$. In addition, we compute the trade imbalance $\eps^{{i}}_t =
n^{i,\mathrm{buy}}_t - n^{i,\mathrm{sell}}_t $, where
$n^{i,\mathrm{buy/sell}}_t$  denotes respectively the number of buyer-
and seller-initiated market orders of
stock $i$ in bin $t$. We choose this proxy for volume imbalance
because the strong fluctuations in the size of the trades are only
moderately compensated by the information that they provide~\cite{jones1994transactions,bouchaud2004fluctuations}. \smallskip

We normalise $\ret^i_t$ and $\eps^i_t$ by their standard deviation
computed over the entire trading period. As
a result, both time series display zero mean and unit variance. This
choice of normalisation has the benefit of making the problem \emph{extensive}
in the following sense: For any linear model that one infers, (such as the one presented in
Section \ref{sec:mult-prop-model}), the results obtained for a larger bin size (say,
one hour) can always be recovered from the results obtained at a finer
scale. Moreover, extensivity allows the predictions of the model not to
depend on the estimation of the local normalization. One does not need 
to build estimators for volatility and volume in the next
five-minutes bin in order to exploit these results.
This would not have been the case had we used a local normalization for
the fluctuations of the returns and the volumes. Still, we have checked
that the choice of a local normalization, while spoiling
extensivity, yields qualitatively similar results.\smallskip

Also note that we have chosen to use real time to measure $t$ as opposed to counting it on a trade-by-trade basis. This is because in the regime of large $N$ that we consider, there would be too many trades, and our dataset would become unmanageable \cite{wang2016cross, wang2016average}.
Finally, the choice of a five-minute bin size allows us to abstract
away from microstructure effects which are not the subject of the present \emph{mesoscopic} study. All along this manuscript time shall be seen as dimensionless, five minutes being the time unit.

%%%%%%%%%%%%%%%%%%%%%%%%%%%%%%%%%%%%%%%%

\section{Market impact and price fluctuations}
\label{sec:market-impact-price}

In this section, we define the multivariate correlation functions relevant to the problem at hand, and investigate their relations.

\subsection{The correlation structure of returns}
\label{sec:return-covar-matr}

The covariance matrix of returns is one of the central objects in
quantitative finance, and is of paramount importance in a number of
applications such as portfolio construction and risk management \cite{markowitz1952portfolio, meucci2009risk}.
Let us recall first some of its most prominent properties.\smallskip

\begin{figure}[t!]
\begin{center}
\includegraphics[width=1\columnwidth]{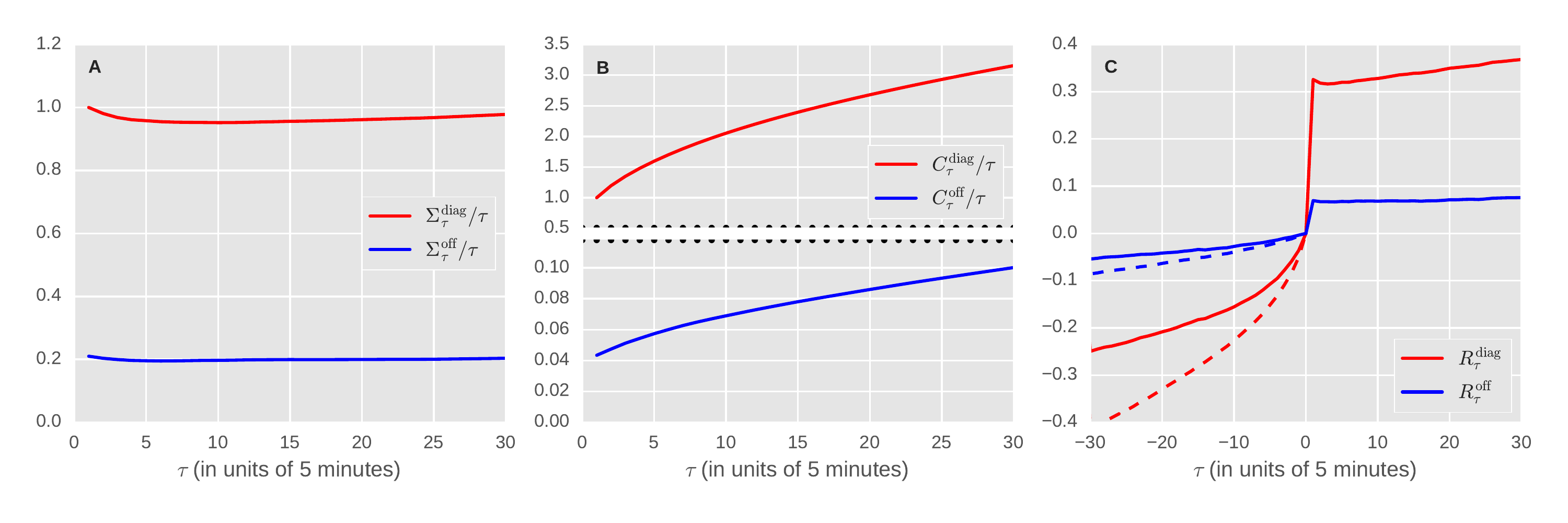} 
\caption{Plots of average diagonal and off-diagonal (a) returns
  covariance (see Eq.~(\ref{retcov})), (b) sign covariance (see
  Eq.~(\ref{signcov})),  and (c) response function (see
  Eq.~(\ref{response})). The dashed lines for the response indicate the
  prediction of the model at negative lags.}\label{mean_covs}
\end{center}
\end{figure}

We denote by $\Cret^{ij}_\tau$ the return covariance of contracts $i$ and $j$ at scale $\tau$, defined as
\begin{eqnarray}
\Cret^{ij}_\tau = \avg[ (\Ret^{{i}}_{t+\tau} - \Ret^{{i}}_t)(\Ret^j_{t+\tau} - \Ret^j_t)]
\label{retcov}
\end{eqnarray}
Figure \ref{mean_covs}(a) displays a plot of the mean diagonal
$\Cret^\diag_\tau = N^{-1}\sum_i \Cret^{ii}_\tau$ and off-diagonal $\Cret^\off_\tau =
(N^2-N)^{-1}\sum_{i\neq j} \Cret^{ij}_\tau$  return covariances rescaled
by $\tau$. As one can see, the diagonal terms of the return covariance
matrix are on average a factor $\sim 5$ larger than the off-diagonal ones.
Microstructural effects
are almost absent in $\Cret^{ij}_\tau$ even at $\tau=1$: we only observe a weak decrease of
the variance at short lags in the signature plot, and the ratio between
covariance and variance --~that determines the so-called Epps effect~\cite{epps1979comovements, toth2009epps}~-- is almost flat in $\tau$. 
This is consistent with the absence of statistical arbitrage price,
because the time scale for these arbitrage effects is nowadays expected to be well below the five-minute
time scale {\cite{toth2006increasing, Large2007, BacryMuzy2014}}.
Finally, one can define the customary return correlation
matrix as
$\Cret^{ij}_\tau(\Cret^{ii}_\tau\Cret^{jj}_\tau)^{-1/2}$.
\smallskip

Figure \ref{cov_matrices}(a) displays a representation of $\Cret^{ij}_\tau$ at $\tau=1$ from which we subtracted its mean ($\approx 0.21$) for better readability, and in which the contracts have been sorted by industrial sector, as indicated by the labels.  As one can see, $\Cret^{ij}_\tau$ displays a strong sectorial structure, in line with previous studies {\cite{laloux1999noise,MantegaLillo, Marsili}}.
The behaviour of the covariance matrix is best understood in its
eigenbasis. Indeed, $\Cret^{ij}_\tau$ is a real symmetric matrix,
so it be diagonalised as
\begin{eqnarray}
\Cret^{ij}_\tau = \sum_a O^{ia}_\tau \Lambda^a_\tau O^{ja}_\tau.
\end{eqnarray}
$O^{ia}_\tau$ is an orthogonal matrix, its columns
correspond to the eigenvectors of $\Cret^{ij}_\tau$, and
$\Lambda^a_\tau$ is a vector made of the corresponding
eigenvalues. Figure \ref{cov_matrices}(b) displays the histogram of
the eigenvalues $\Lambda^a_\tau$ at $\tau=1$. We have assessed their
stability by verifying that $\Lambda^a_\tau \propto \tau$, as it was the
case for the average quantities displayed in
Fig.~\ref{mean_covs}(a).
Interestingly, we find the eigenvectors $O^{ia}$ to be stable in time, indicating that
the directional structure of the market
is consistent across scales ranging from some minutes to one day,
while its associated fluctuations increase linearly.\footnote{This however does not mean that there is no intraday seasonality in the correlation 
structure, see~{\cite{RomainAllez}}.}
The value of the largest eigenvalue
$\Lambda^0_1 \approx 62$, indicates that $\Lambda^0_1/N \approx 23\%$ of the total
variance of the system can be explained by this mode, in good agreement
with Ref. \cite{hasbrouck2001common}. Often referred to as the market mode,
it corresponds to a collective --~and rather homogeneous~-- mode, as can be seen
on Fig.~\ref{cov_matrices}(c). The
next few modes after the market mode, individually, explain a considerably smaller part of the
variance. Their structure supports
an economic interpretation in terms of industrial sectors (see
Fig.~\ref{cov_matrices}(c) and Ref. \cite{laloux1999noise}). The subsequent modes fall into a noise band that is
roughly described by a Mar{\v{c}}enko-Pastur distribution \cite{marvcenko1967distribution,bouchaud2009financial} (see red
curve on Fig.~\ref{cov_matrices}(b)), due to the fact that the number
of stocks is of the same order of magnitude as the number of
observations, making it impossible to obtain a statistically accurate estimation of
all the modes.

\subsection{The correlation structure of the trade signs}
\label{sec:sign-covar-matr}
In order to investigate the relation between returns and trade sign
imbalance, it is natural to define a covariance matrix for the signs,
and to compare its structure with the one built out of the returns.
Accordingly, we define the lagged
covariance of signs as
\begin{eqnarray}
\csgn^{ij}_\tau &=& \avg[\eps^{{i}}_{t+\tau}  \eps^j_t] \ .
\end{eqnarray}

Its behaviour is radically different from that of returns.
While returns are uncorrelated ($\Cret^{ij}_\tau\sim \tau$ after a few
trades) compatible with statistical efficiency of prices, signs are well known to be long-range
correlated, as $\csgn^{ij}_\tau \sim \tau^{-\gamma}$ with
$\gamma \sim 0.5$ (see Appendix). This result stems from
the fact that in limit order markets investors split their trading decisions
into smaller pieces in order to avoid excessive costs, because instantaneously available liquidity
at the best quotes is small \cite{lillo2004long}, much smaller than the daily volume. This yields the famous
anomalous response puzzle \cite{bouchaud2004fluctuations}: Prices diffusive despite being driven by
trades which themselves are superdiffusive.

A well-known solution to this
problem is that of the linear propagator model (or, equivalently, the surprise model),
postulating that trades in the most probable direction impact
the price less than those in the unexpected one
\cite{bouchaud2004fluctuations,lillo2004long,bouchaud2006random,bouchaud2008markets}. While
this model has been thoroughly explored in one dimension (with extensions to multi-order types, \cite{eisler2012price,eisler2011models,taranto2016linear1}), its richer multi-dimensional counterpart has not been fully considered yet. A multivariate framework allows to precisely formulate a number of
questions that are central to our study, and that cannot be addressed
in a one-dimensional setting. What is the role of the trade sign
process in shaping the cross-sectional structure of the return
correlations? Is there such a thing as a market mode for signs (our
proxy for liquidity)? Are there liquidity sectors?
\begin{figure}[t!]
\begin{center}
\includegraphics[width=1\columnwidth]{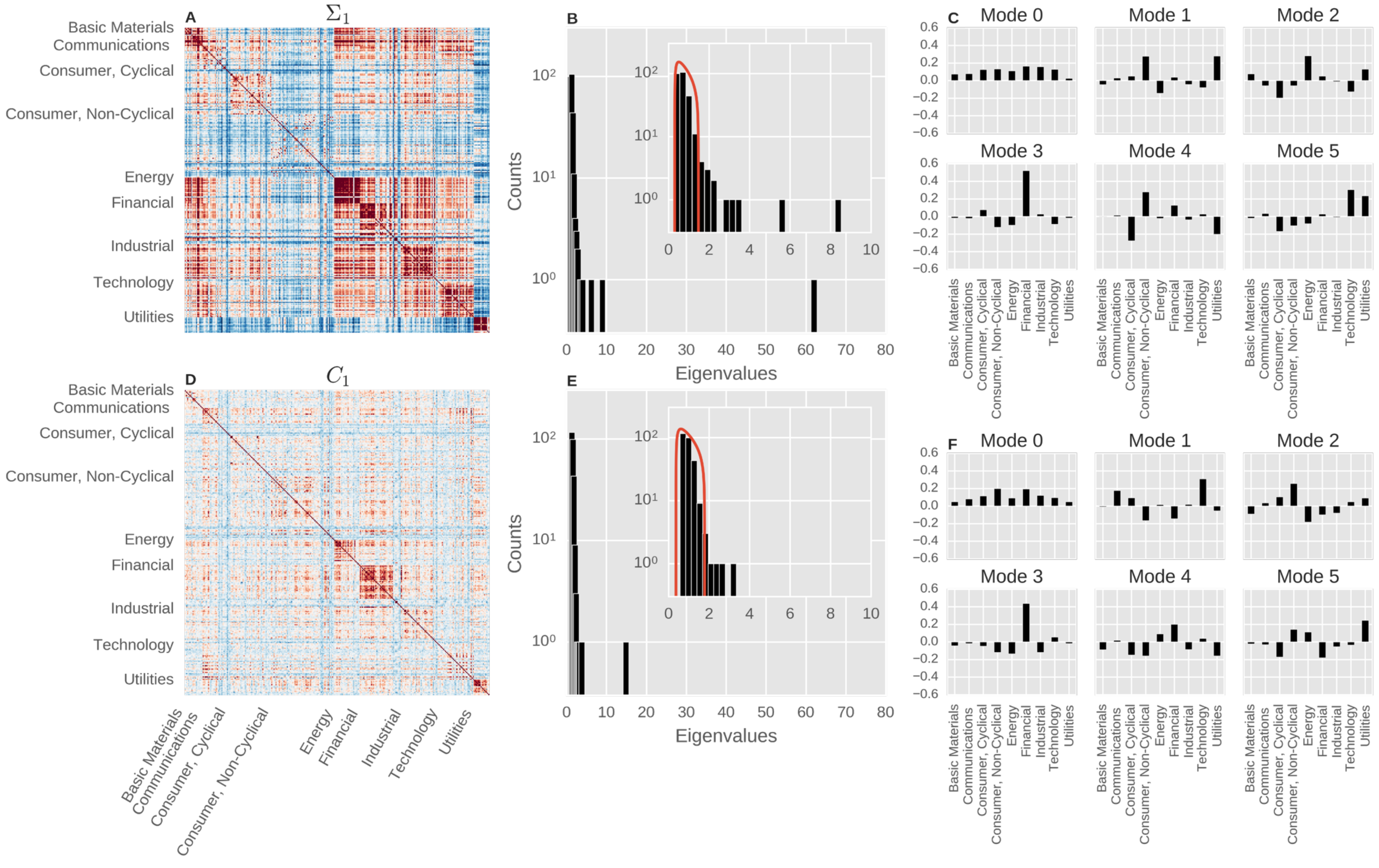}
\caption{(a) Plot of the returns covariance matrix $\Cret^{ij}_\tau$ at lag $\tau=1$. (b) Histogram of eigenvalues of $\Cret^{ij}_1$. (c) Composition of the eigenvectors (weights per sector). (d-f) Same plots for the sign covariance matrix $\Csgn^{ij}_\tau$. }\label{cov_matrices}
\end{center}
\end{figure}
In order to  push this parallel further, it is useful to define the 
equal-time covariance $\Csgn^{ij}_\tau$ of the cumulated trade sign process, which is analogous to $ \Cret^{ij}_\tau$, defined as
\begin{eqnarray}
\Csgn^{ij}_\tau = \avg[(\Eps^{{i}}_{t+\tau} - \Eps^{{i}}_t)(\Eps^j_{t+\tau}-\Eps^j_t)] \ ,
\label{signcov}
\end{eqnarray}
where the $\eps^{i}_t$ are the analogue of the ``returns'' for $\Eps^i_t$: $\eps^{i}_t = \Eps^i_t - \Eps^i_{t-1}$.
Figure \ref{mean_covs}(b) displays a plot of the mean diagonal $C^\diag_\tau$ and off-diagonal $C^\off_\tau$  sign covariances rescaled by $\tau$. Similarly to $\Cret^{ii}_\tau$, the diagonal terms are on average larger than the off-diagonal ones, only this time by a factor $\sim 30$. After a short sublinear regime,
the results show superlinear time dependence at large $t$, consistent
with the long-range correlation of signs for a single asset. Figure
\ref{cov_matrices} shows that, in contrast with the covariance of
prices, the covariance of signs displays no or very weak sectorial
structure.
 Although
the first mode of the sign covariance also corresponds to a market
mode (delocalized and rather homogeneous), it is weaker.
Additionally, one has a small number of ``sectorial'' modes
out of the noise band \cite{bouchaud2007large}, that even in this case are coherent in time,
showing a time-overlap close to~1. 
Despite this, only the market mode is aligned with the market mode of returns.
All the other modes show surprisingly small overlap with their
return counterparts (see Fig.~\ref{overlap}(b) for a quantitative discussion on the fraction of common modes).
\smallskip

\medskip

\subsection{Price response}
\label{sec:price-response}

Do trades shape the return covariance matrix? Or does it result from other
mechanisms such such as quote revisions, that do not involve trading volume?
In order to address such questions, one needs to look into yet
another quantity, the market response $\resp^{ij}_\tau$ defined as:
\begin{eqnarray}
\resp^{ij}_\tau = \avg[ (\Ret^{{i}}_{t+\tau} - \Ret^{{i}}_t) \eps^j_t] \ .
\label{response}
\end{eqnarray}
This measures the average price change of contract $i$ at time $t+\tau$, after
experiencing a sign imbalance $\eps^j_t$ in contract $j$ at time $t$. Figure
\ref{mean_covs}(b) displays a plot
of the mean diagonal $R^\diag_\tau$ and off-diagonal $R^\off_\tau$
responses. The diagonal terms are on average larger than the
off-diagonal ones by a factor $\sim 5$. This is consistent with the ratio of
the corresponding diagonal/off-diagonal factors for the price and sign
covariances, and with the results of Refs. \cite{wang2016average, wang2016cross}. The response at positive times is roughly
constant, consistently with the hypothesis of a statistically efficient price.
In other words, the current sign does not predict future returns. The
behavior at negative lag  indicates that the current return allows some prediction of the sign imbalance, an effect that has been extensively
investigated in Refs. \cite{taranto2016linear1, taranto2016linear2}.\footnote{We will disregard in the
following the behavior of returns at negative lags, and only focus
on the positive part of the curve, that is equivalent to assuming no
price-sign correlation, that is approximately correct for small tick
stocks, and breaks down at high frequency and for large tick stocks due to microstructural
effects \cite{taranto2016linear1, taranto2016linear2, dayri2015large}.} It is worth mentioning that, other than the expected amplitude difference, the off-diagonal response shows the same temporal behaviour than its diagonal counterpart.

\medskip

%%%%%%%%%%%%%%%%%%%%%%%%%%%%%%%%%%%%%%%%%%%%%%

\section{A simple model for cross-impact}
\label{sec:simple-model}
In this section, we present and analyse the implications of the multivariate propagator model, which shall
allow us to explain within a coherent framework the stylised facts
discussed above.

\subsection{The multivariate propagator model}
\label{sec:mult-prop-model}

As we shall see the simplest linear model $(i)$ describing the
cross-sectional structure of covariance matrices, $(ii)$ accounting for
their dynamical structure, and $(iii)$ assuming future signs are weakly affected by
recent past returns, is the multivariate \emph{propagator model}:
\begin{eqnarray}
\Ret^i_t &=& {\Ret^i_0 +  \sum_j \sum_{t'=1}^t } \Prop^{ij}_{t-t'} \eps^j_{t'} + W^{{i}}_{t} \ . \label{def_G}
\end{eqnarray}
This expresses the price variations
of contract $i$ as a linear regression on the past sign imbalances
of all assets $j$. The matrix $\Prop^{ij}_\tau$ is customarily called the propagator, as it
describes the effect of the trade sign imbalance of contract $j$ at time $t$ on
the price of contract $i$ at time $t+\tau$\footnote{Note that the model is \emph{self-consistent}, in the sense that artificially
splitting the same contract $i$ in two fully correlated
instruments $i_1$ and $i_2$ yields a completely equivalent dynamics
for the returns $\Ret^{i_1}_t = \Ret^{i_2}_t$
under any transformation of the type $\eps^i = \eps^{i_1} + \eps^{i_2}$, provided
that $\Prop^{i_1 i_1} = \Prop^{i_1 i_2} = \Prop^{i_2 i_1} = \Prop^{i_2 i_2}$, 
$\Prop^{i_1j} = \Prop^{i_2 j}$ and $\Prop^{j i_1} = \Prop^{j i_2}$ for all $j$. This is due to our choice of extensive
units for the volume. Due to our requirement of unit variance for
  the series of $\ret^i_t$ and $\eps^i_t$, in order to obtain
  consistency one obviously has to reintegrate units back into the
  problem. We believe this self-consistency condition to be a
necessary requirement for any  satisfactory model for cross-impact.}. The quantities $W^{{i}}_t$ are
defined by $w^{{i}}_t = W^{{i}}_t - W^i_{t-1}$, where the $w^{{i}}_t$ are i.i.d.\
idiosyncratic noises with zero mean and covariance matrix given by
\begin{eqnarray}
  \label{eq:2}
  \avg[w^{{i}}_t w^j_{t'}] = \cret_W^{{ij}} \delta_{t-t'} \, ,
\end{eqnarray}
so that the covariance of the process $W^{{i}}_t$ is linear in time, and is given by
\begin{eqnarray}
\Cret^{ij}_{W,\tau} &=& \avg[(W^{{i}}_{t+\tau} - W^{{i}}_t)(W^j_{t+\tau} - W^j_t)] 
= \cret_W^{{ij}} \tau \ .
\end{eqnarray}
{ 
Since we consider a setting in which $\varepsilon_t$ is a stationary process, and both $G_\tau$ and the correlations of $\varepsilon_t$ decay to zero at large lags, it's straightforward to check that the model defined by~(\ref{def_G}) converges to a stationary state at large times. Accordingly, in the main text will always refer to the value of the observables $C$, $\Sigma$ and $R$ computed under the stationary measure of the process $\avg[\cdots]$. In the calibration of the process we will also assume stationarity to hold, by imposing time-translational invariance for the correlations of $\varepsilon_t$  (see Appendix).}
\smallskip
  
\begin{figure}[t!]
\begin{center}
\includegraphics[width=11cm]{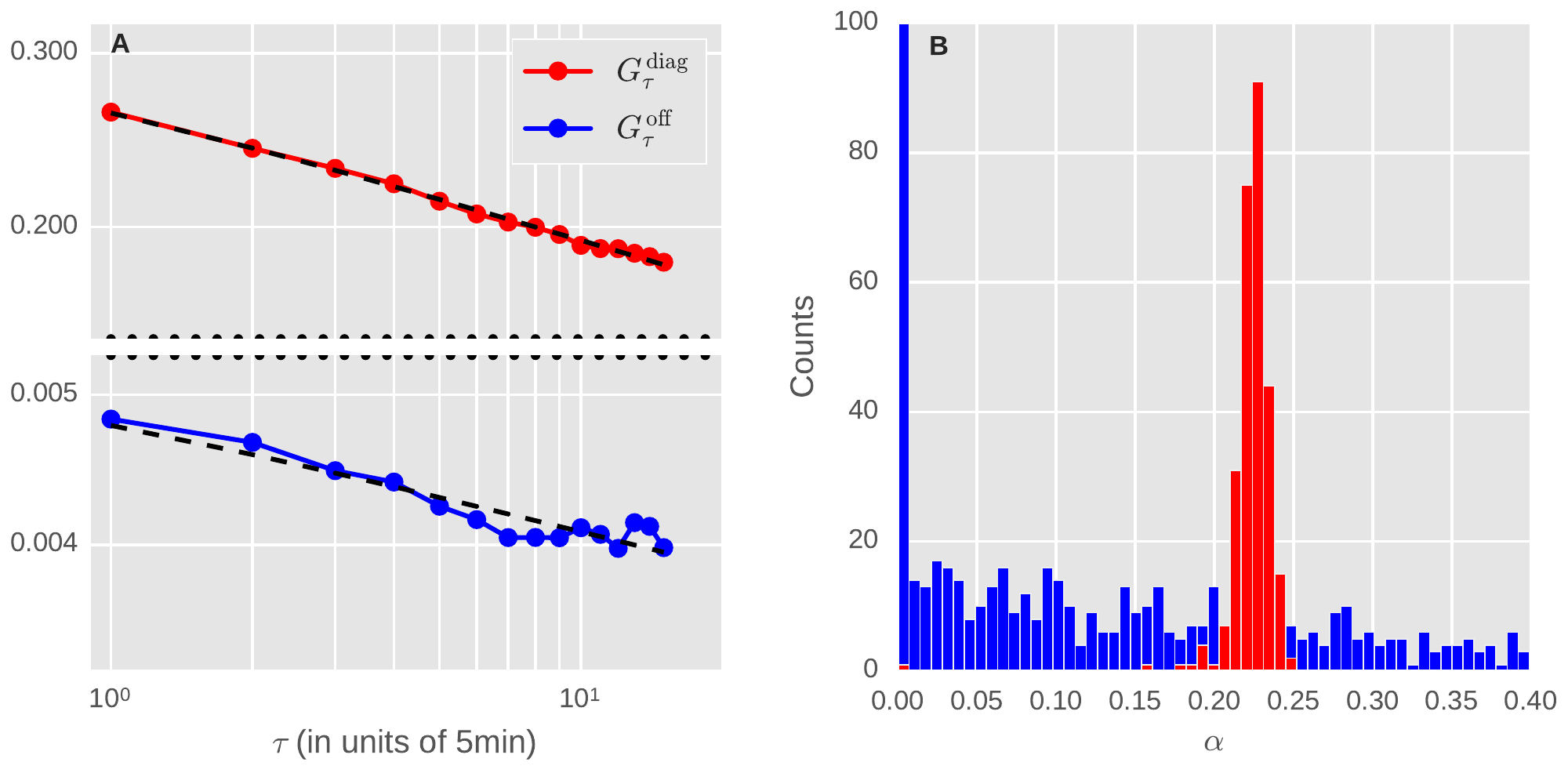}\caption{(a) Plot of the mean diagonal and off-diagonal propagators. (b) Corresponding histogram of fitted slopes $\beta$, as given by Eq.~(\ref{time_dep}). } \label{means_G}
\end{center}
\end{figure}

We have fitted the propagator matrix $\Prop^{ij}_\tau$
from data. Figure~\ref{means_G}(a) displays a plot of the mean diagonal
$\Prop^\diag_\tau$ and off-diagonal $\Prop^\off_\tau$ propagators that we
have obtained under a non-parametric inversion of the model. [See also Section~\ref{sec:dimensional_reduction} for a
comparison of the different inversion techniques that we have adopted.]
The diagonal terms are on average larger than the off-diagonal ones by a factor
$\sim 50$, see Figure~\ref{means_G}(a). Both are consistent with
a power-law decay in time, as expected from the one-dimensional case.
Figure~\ref{means_G}(b) shows fluctuations in the plot, while the slope of the diagonal components is rather well defined, that of the off-diagonal presents large fluctuations. However such fluctuations average away, as they seem to be structureless.
More precisely, despite the large difference in
magnitude between the diagonal and the off-diagonal entries of $\Prop^{ij}_\tau$,
they are both compatible with a power-law decay:
\begin{equation}
  \label{eq:factorized}
\Prop^{ij}_\tau = \Prop^{ij} \left(1+\frac{\tau}{\tau_0}\right)^{-\beta}.
\label{time_dep}
\end{equation}

\begin{figure}[t!]
\begin{center}
\includegraphics[width=16cm]{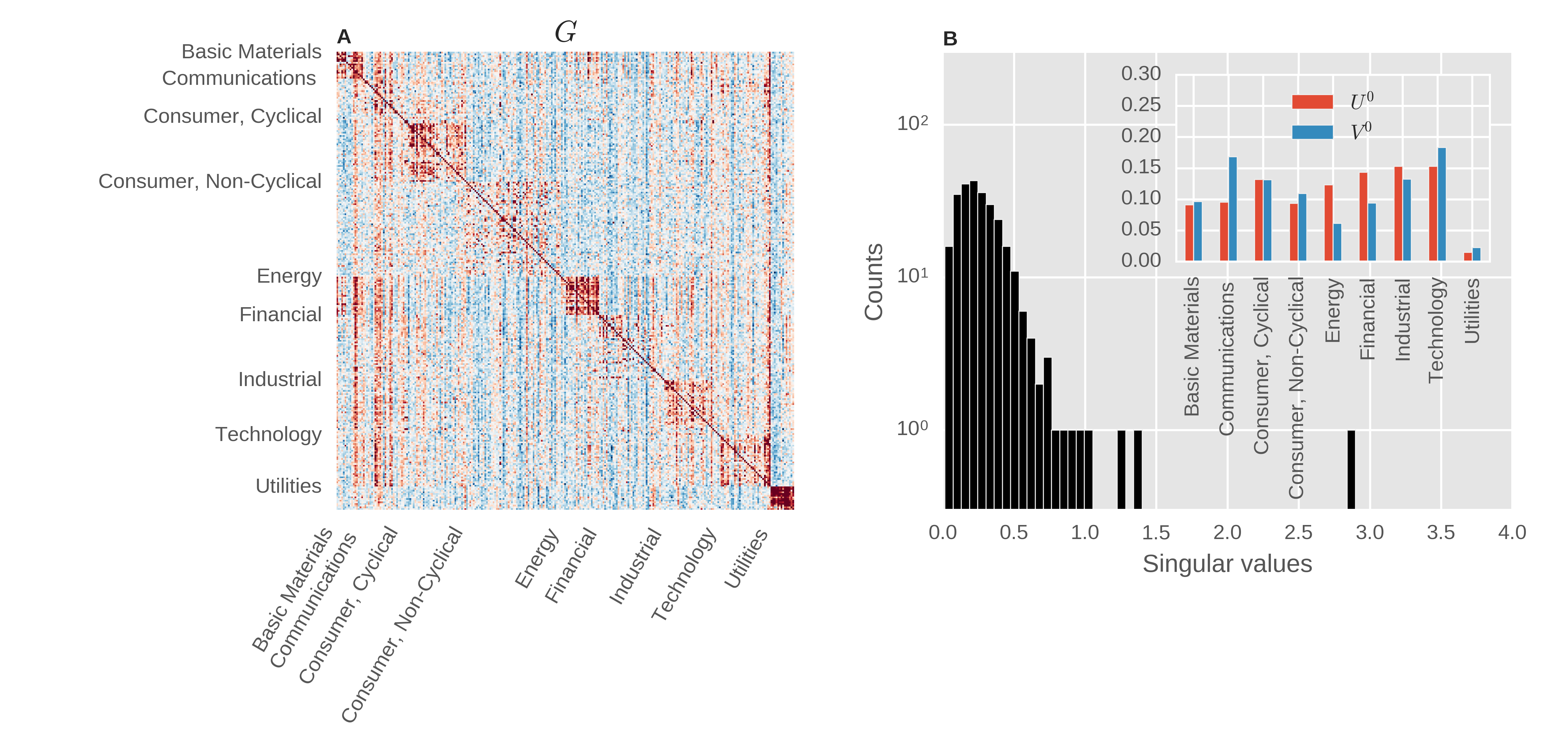} \caption{ (a) Plot the propagator matrix $G^{ij}$ as obtained from the factorised scheme.  (b) Histogram of singular values and composition of the ground singular vectors.} \label{G_matrix}
\end{center}
\end{figure}

This constitutes a \emph{factorized model} in which the temporal and
cross-sectional parts are separated, and as we shall see this will
facilitate the analysis by reducing the dimensionality of the problem.
Fitting Eq.~(\ref{eq:factorized}) to the
diagonal and off-diagonal data yields $\beta^{\mathrm{diag}} = 0.14$,
$\beta^{\mathrm{off}} = 0.09$, $\tau_0^{\mathrm{diag}}=0.30$ and
$\tau_0^{\mathrm{off}}=0.32$. Figure \ref{G_matrix}(a) displays a plot
of $\Prop^{ij}$ from which we subtracted its mean for better readability,
and in which the stocks have been sorted by industrial sector, as
indicated by the labels.  As one can see,  $\Prop^{ij}$
displays a stronger sectorial structure\footnote{Also note the presence of vertical stripes in Figure
\ref{G_matrix}(b), indicating that --~while the choice of the standard
deviation of returns for normalizing returns allows to obtain a
homogeneous rows~-- using the standard deviation at $t=1$ for the
signs is not the best choice to obtain a uniform $\Prop^{ij}$. Of course,
this feature can be reabsorbed through a suitable
definition of the units of $\eps^i_t$.} than $C^{ij}_t$.

In order to address this issue more quantitatively, we introduce the
singular-value decomposition of $\Prop^{ij}$, defined as
\begin{eqnarray}
\Prop^{ij} = \sum_a U^{ia} S^a V^{ja}.
\end{eqnarray}
$U^{ia}$ and $V^{ia}$ are real orthogonal matrices, the columns of which
correspond to the left/right singular vectors of $\Prop^{ij}$, and where
$S^a$ is a vector made of the corresponding singular values.
The interpretation of the decomposition is straightforward: For a given
$a$ the
value $S^a$ is the increase of a linear combination $U^{ia}$ of stock prices
after the combination of trades $V^{ia}$. Figure
\ref{G_matrix}(b) displays the histogram of singular values
$S^a$. 
Fig. \ref{G_matrix}(b) shows, among other things, that a
market-neutral net imbalance has a smaller impact on prices than a
directional one. In fact, due to $U^{i0} \approx V^{i0} \approx
N^{-1/2}$ (see the inset of Fig. \ref{G_matrix}(b)), trading one
standard deviation of the imbalance of the market mode costs roughly three standard
deviations of its price, while all the other modes have a consistently smaller impact.

\subsection{Response function and price covariation}
\label{sec:price_cov_decl}

{
Having found the propagator, 
we can 
investigate the interplay of  $\Prop^{ij}_\tau$ with $\Csgn^{ij}_\tau$ in shaping the response function and 
return correlation.
In particular, within the propagator model one finds:
\begin{eqnarray}
R^{ij}_{\tau} &=& \sum_{k} \Big[ \sum_{\tau' = 0}^{\tau-1} \Prop^{ik}_{\tau'} \csgn^{kj}_{\tau'-\tau}   +\sum_{\tau' = \tau}^{\infty} \left( \Prop^{ik}_{\tau'}-\Prop^{ik}_{\tau'-\tau} \right)\csgn^{kj}_{\tau'-\tau}  \Big]  \ , \label{resp_temp}
\end{eqnarray}
and:
\begin{eqnarray}
\label{eq:decomposition}
\Cret^{ij}_\tau &=&  \Cret^{ij}_{G,\tau} + \Cret^{ij}_{W,\tau} \ ,
\end{eqnarray}
where:
\begin{eqnarray}
\Cret^{ij}_{G,\tau} &=& \sum_{k,l} \Big[ \sum_{\tau',\tau'' = 0}^{\tau-1} \Prop^{ik}_{\tau'} \csgn^{kl}_{\tau'-\tau''} \Prop^{jl}_{\tau''}  + 2\sum_{\tau'= 0}^{\tau-1} \sum_{\tau'' = \tau}^{\infty} \Prop^{ik}_{\tau'} \csgn^{kl}_{\tau'-\tau''}   \Big( \Prop^{jl}_{\tau''} -  \Prop^{jl}_{\tau''-\tau} \Big) \nonumber \\
&&  \ \ \ \ \  \ \ \ \ \  \ \ \ \ \ \ \ \ + \sum_{\tau',\tau'' = \tau}^{\infty} \Big( \Prop^{ik}_{\tau'} -  \Prop^{ik}_{\tau'-\tau} \Big) \csgn^{kl}_{\tau'-\tau''}   \Big( \Prop^{jl}_{\tau''} -  \Prop^{jl}_{\tau''-\tau} \Big)
\Big]  \ . \label{cov_temp}
\end{eqnarray}
}

\begin{figure}[t!]
\begin{center}
\includegraphics[width=9cm]{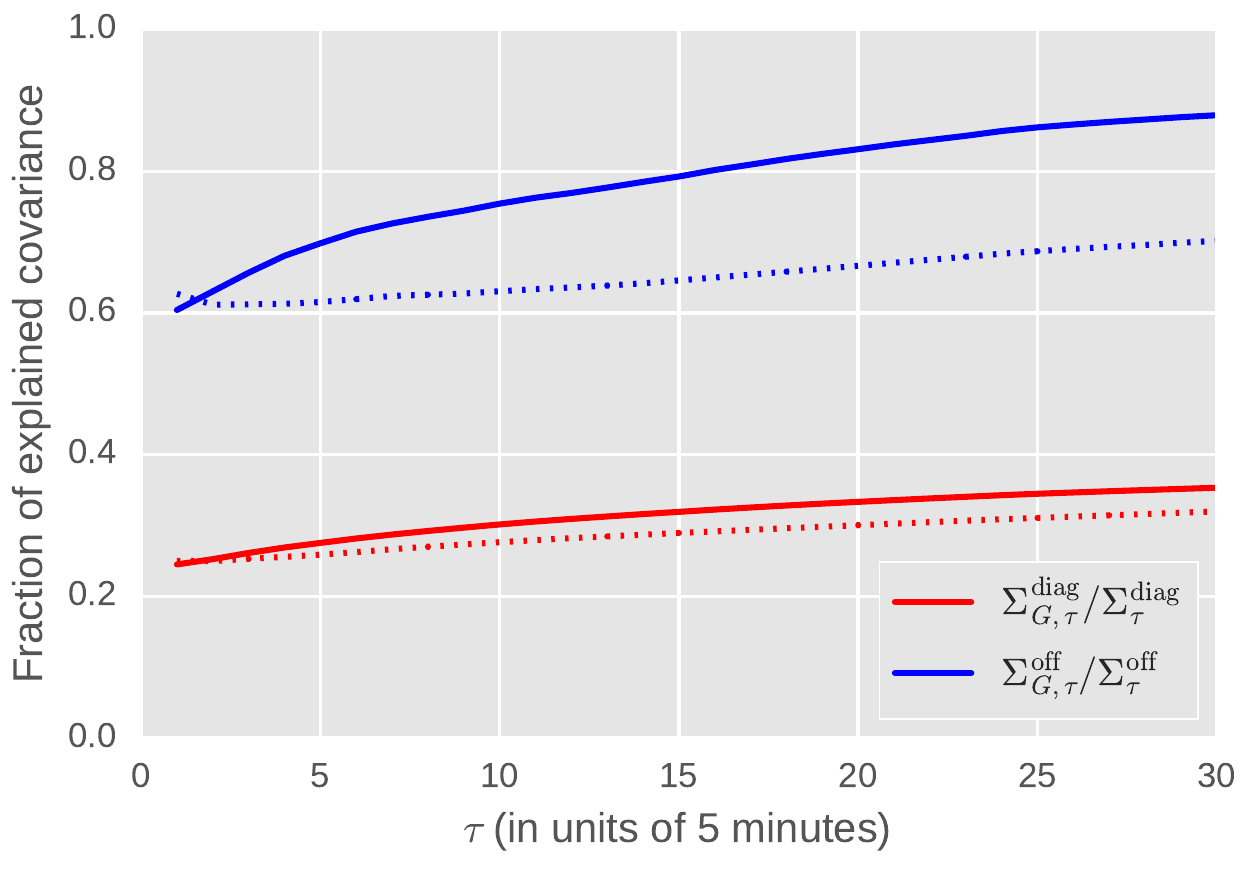} \caption{ {Plot of the fraction of explained diagonal and off-diagonal covariance as given by Eq.~(\ref{eq:decomposition}) as a function of the lag. The solid lines were obtained by extrapolating the sign correlation to infinity while the dotted lines are the result of truncating the past to a maximum lag equal to $T=30$.}  } \label{explained_cov_fig}
\end{center}
\end{figure}

\begin{figure}[h!]
\begin{center}
\includegraphics[width=\columnwidth]{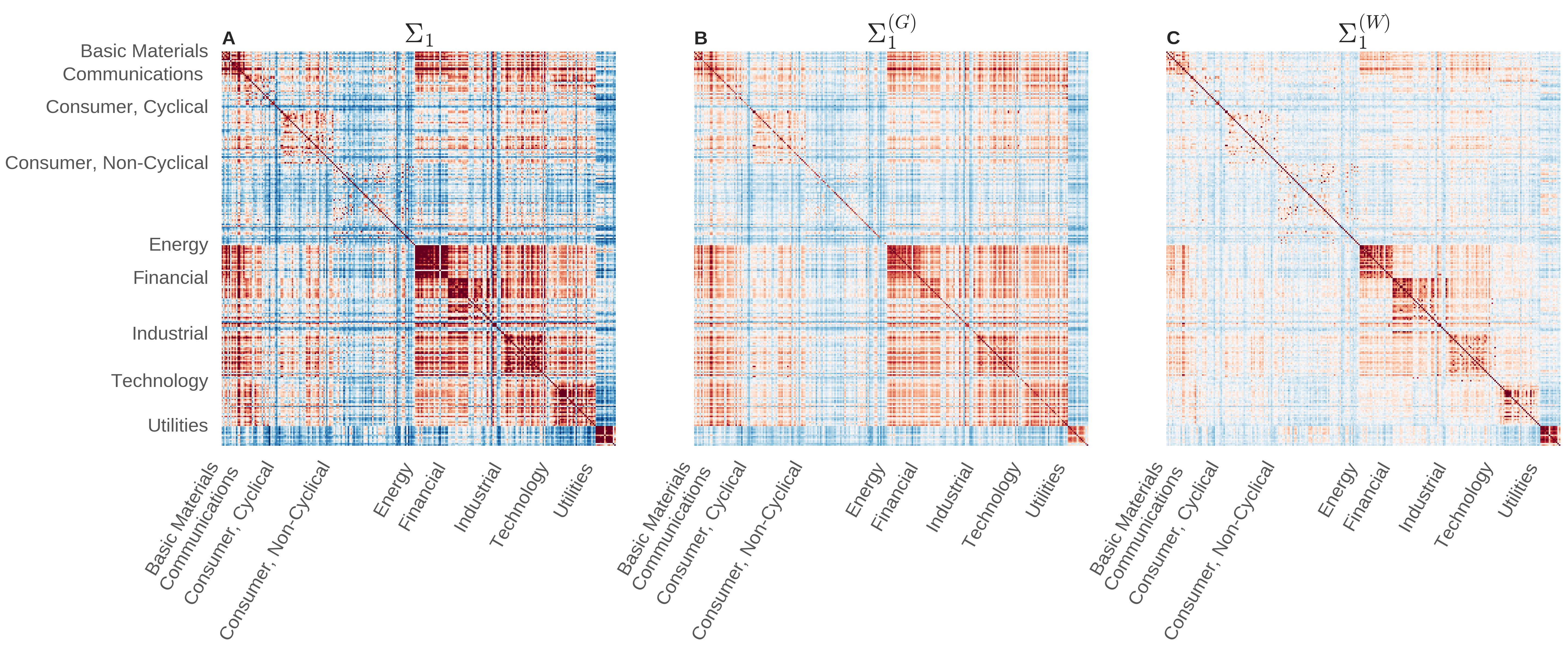} \caption{Plot of (a) the returns covariance matrix, (b) $\Cret^{ij}_{G,\tau}$,  and (c) $\Cret^{ij}_{W,\tau}$,  at lag $\tau=1$ (see Eq.~(\ref{eq:decomposition})). Note that the matrices having been substracted of their mean for better visibility of their structure. } \label{CvsGCG}
\end{center}
\end{figure}

This is an extension of the result found in Refs. \cite{bouchaud2004fluctuations,pasquariello2013strategic}
in a linear equilibrium setting. The time-behavior of the first term $\Cret^{ij}_{G,\tau} $ captures the
dynamics of the model, that is very similar to the one 
found in the one-dimensional model. In that case, even if at large
times $\csgn_\tau \sim \tau^{-\gamma}$  with
$\gamma \approx 0.5$, the long range dependence of the resulting
propagator $\Prop_\tau \sim \tau^{-\beta}$ with $\beta \approx 0.25$ is able
to compensate the long-range dependence of the imbalances and restore
the diffusivity of price, which indeed requires $\beta=(1-\gamma)/2$~\cite{bouchaud2004fluctuations,bouchaud2008markets}. In this more general setting, as the time
behavior of the model is found to be well-described by the factorized
model~(\ref{eq:factorized}), we are offered the same solution to
reconcile the behavior of sign and return correlations.
With these definitions, $\Cret^{ij}_{G,\tau} $ denotes the part of the return covariance explained
by the impact of transactions, while $\Cret^{ij}_{W,\tau} $ stands for its unexplained
component, for example due to news.
Figure \ref{explained_cov_fig} displays the fraction of explained diagonal and off-diagonal covariance, which appears to increase with   the lag. Interestingly, one can see that while only $\approx 25$-$35\%$ of the diagonal
variance can be explained by impact,\footnote{{Note that this number is significantly smaller than the 60-70$\%$ fraction quoted in~\cite{eisler2012price,eisler2011models} is related to low frequency nature of the 5-minute binned data used in the present study.}} this figure rises to
$\approx 60$-$90 \%$ for its off-diagonal counterpart, meaning that the
propagator model is somewhat more efficient to explain the covariance
than it is to account for the variance. 
 It is also
interesting to mention that the propagator model is successful in reproducing
the sectorial structure of the covariance matrix. 
For a visual interpretation, Fig.~\ref{CvsGCG} displays plots of the
three matrices that appear in Eq.~(\ref{eq:decomposition}). 
\smallskip

In order to assess whether the propagator model helps in understanding
the direction of the risk modes of the market (and in particular, the
composition of the sectors), we 
raise the following question: does $\Cret^{ij}_{G,\tau}$ explain more
of the price covariance structure than the sign covariance $C_\tau$ alone?  To answer this
quantitatively we compare the overlap of the eigenvectors of
$\Cret_\tau $ with those of $\Cret_{G,\tau}$ and with those of
$C_\tau$. 

\begin{figure}[h!]
\begin{center}
\includegraphics[width=1\columnwidth]{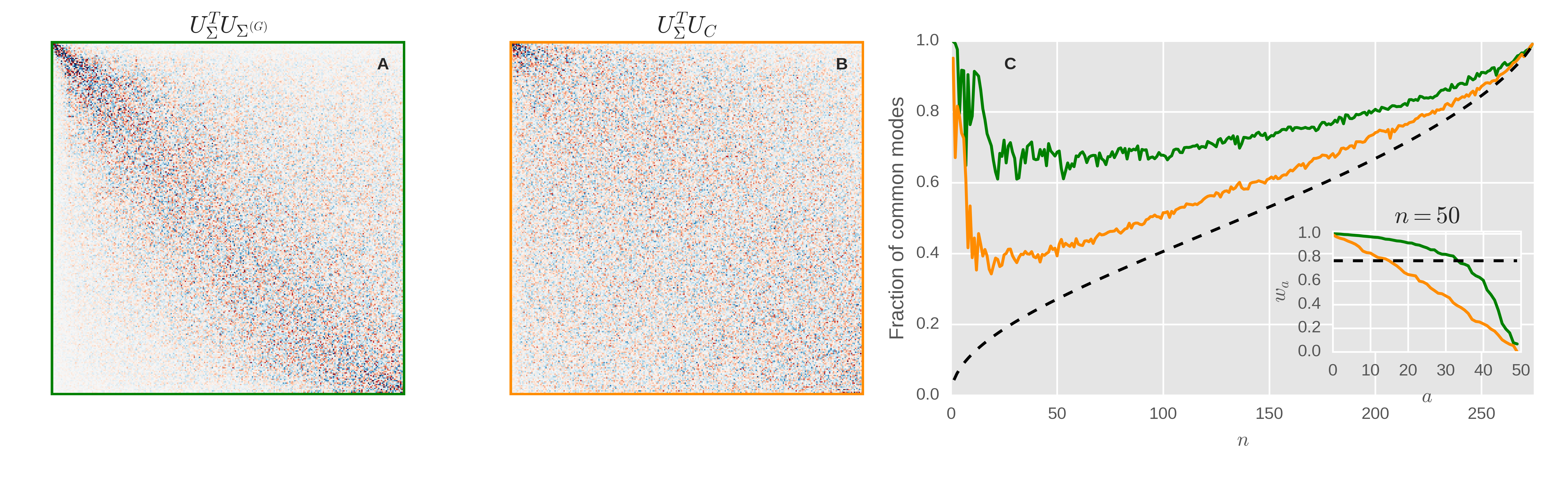} \caption{Plot of the overlap of eigen-rotation matrices of the returns covariance matrix with (a) $\Cret^{ij}_{G,\tau}$ (see Eq.~(\ref{eq:decomposition})), and (b) the sign covariance matrix. (c) Plot of the fraction of common modes as defined in the text.} \label{overlap}
\end{center}
\end{figure}

More precisely, if we denote the eigenvectors of these
matrices by $U_{\Sigma},U_{C}$ and $U_{\Sigma_G}$, we have
computed the overlap matrices $U^T_{\Sigma}U_{C}$ (see
Figure~\ref{overlap}(a)) and $U^T_{\Sigma}U_{\Sigma_G}$
(see Figure~\ref{overlap}(b)). As one can 
see with the naked eye, the eigenmodes of $\Cret_\tau$ have
significantly larger overlap with $\Cret_{G,\tau}$ than there is with $C_\tau$.
Figure~\ref{overlap}(c) displays a plot that captures quantitatively the latter
statement and is constructed as follows: $(i)$ we crop each of the
overlap matrices at $n\in \llbracket1,N\rrbracket $, $(ii)$ compute
their singular values $\{w_a\}_{a\in \llbracket1,n\rrbracket}$ and
sort them in decreasing order (the inset shows the singular value
spectra at $n=50$), and $(iii)$ compute the so-called fraction of
common modes $\left(\prod_{a=1}^{n} w_a \right)^{1/n}$ and plot it as
a function of $n$. The dashed black line represents the theoretical
expectation for the noise level \cite{bouchaud2007large}.
This measure represents the volume of the common subspace spanned by
the first $n$ eigenvectors, and is a very strict measure of
similarity, which is why the results in Figure~\ref{overlap}(c) are
rather remarkable: one sees that the directional structure of the return covariance
matrix can be predicted very well  using liquidity variables only.

\subsection{Direct and cross-impact}

A lot of the covariance and part of the variance come from impact, but how
to measure the direct influence of impact versus its cross-sectional component?
For the response  $\resp^{ij}_\tau$  {and covariance $\Cret^{ij}_{G,\tau}$}, one can simply use the following relations, {which we have written in a diagrammatic way for the sake of readability. Note that the time structure has been omitted but can be easily recovered as each of the following terms has the temporal structure  given in  Eqs.~(\ref{resp_temp}) and (\ref{cov_temp}). Red and blue filled circle signify returns and signs respectively. Empty circles imply exclusive sum over the products. Solid arrows represent propagators and dashed lines stand for sign correlations. 
}

%%%%%%%%%%%%%%%%%% Diagrams
\begin{figure}[h!]
\begin{center}
  \hspace{-1.1em}
\includegraphics[width=16cm]{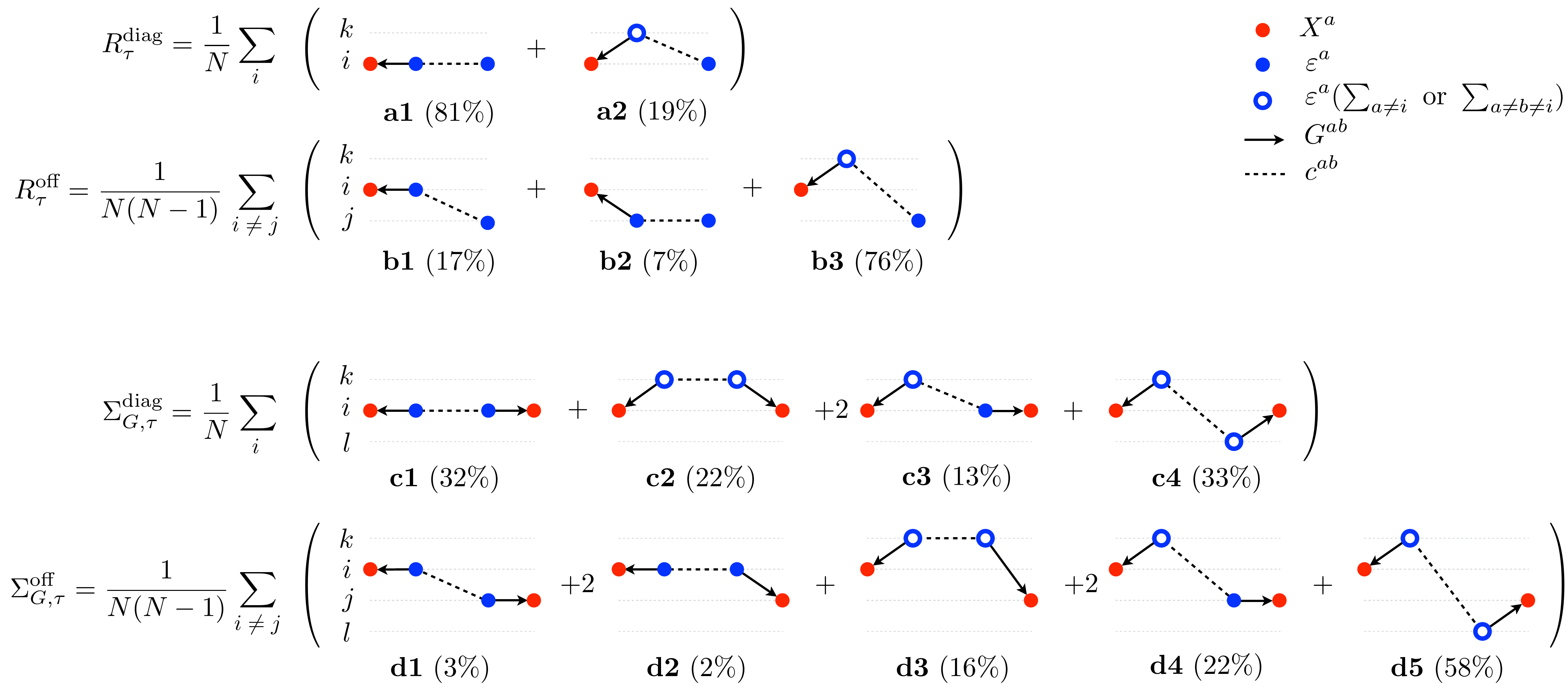}
\label{diagramatic}
\end{center}
  \vspace{-2.5em}
\end{figure}

\noindent Below each term, we have indicated its relative contribution to the
average self/cross-response/covariance.
The interpretation of each term in the response is as follows:
\begin{description}
\item[\textbf{a1}] 
\emph{Self-response via direct impact}. This is the classical term considered in most works on impact: trading in product $i$ impacts the price of $i$ itself.
\item[\textbf{a2}] \emph{Self-response mediated by cross-trading and
  cross-impact}. This term is induced by the order flow on all the stocks $k$
  that are correlated to $i$. This causes market makers to
  include this extra information in their price for $i$.
\item[\textbf{b1}] \emph{Cross-response mediated by cross-trading and
  direct impact}. The mechanism is similar to \textbf{a1},
  except that the order flow on $j$ now induces an imbalance on $i$,
  that translates into a price change via direct impact. 
  \item[\textbf{b2}] \emph{Cross-response mediated by direct trading and
  cross-impact}. Here the market markers react to the order flow on $j$
  by updating their quotes on product $i$.
\item[\textbf{b3}] \emph{Cross-response market mediated by
    cross-trading and cross-impact}. Trading in a stock $j$ that is
  correlated with a large number of other stocks $k$. The market maker
  observes the order flow all of those, and adjusts his quote of $i$
  based on this aggregate information.
\end{description}
Regarding the average weights of the different terms, it is
interesting to notice that while most of the self-response can be
explained through direct impact, this is no longer the case for the
cross-response. For the latter, the dominating mechanism is
\textbf{b3}, implying that most of the cross-response is mediated
by \emph{delocalized modes} (such as the market mode, or large sectors). This is one of the central message of this paper.
{Note that the same story can be told for the returns covariance with similar conclusions for the off-diagonal contribution.}

\subsection{Finite size scaling}
\label{sec:finite-size-scaling}

As the  main goal of this study is the characterization of the interactions among a large number of
stocks, the fact that we only consider a sample of 275 instruments (whereas the US stock market amounts to several thousands of them), might seem restrictive.
Such a relatively smaller sample implies that the order flow for
all those missing products is -- to us -- unobserved, even though the
interaction ($C$ and~$G$) between stocks is on average positive. This may therefore lead
to an overestimation of the magnitude of the propagators, which would
then depend on system size.

\begin{figure}[t!]
\begin{center}
\includegraphics[width=9cm]{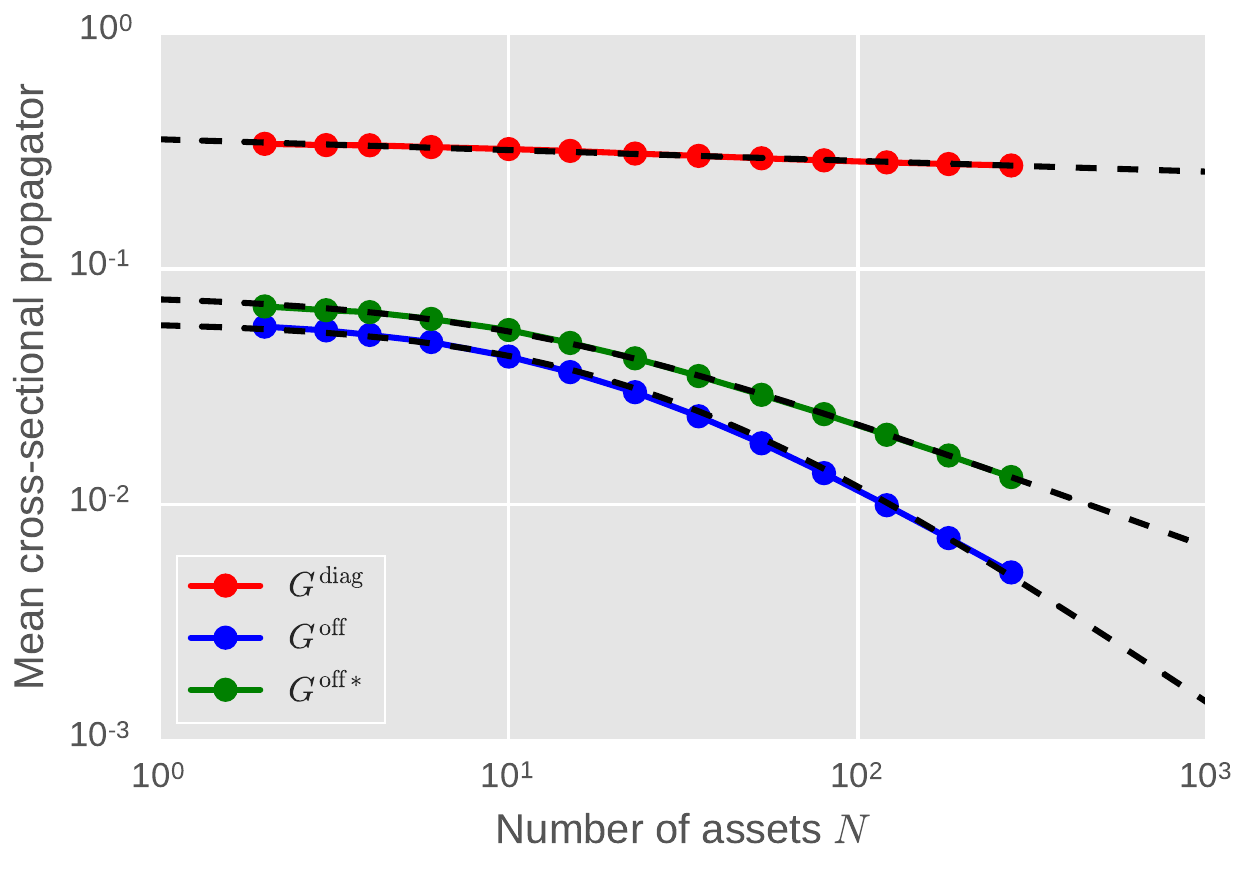} \caption{Plot of the mean diagonal ($G^\mathrm{diag}$) and off-diagonal ($G^\mathrm{off}$) propagators, as well as the mean of the off-diagonal elements where the traded and the impacted stock belong to the same sector ($G^{\mathrm{off}*}$). All curves are computed for $N$ stocks, as a function of $N$, and each data point results from the average of $10^3$ random bootstrap subsamples for each of which we perform cross-sectional propagator inversions.}
\label{scaling}
\end{center}
\end{figure}

In order to empirically verify this effect, we have fitted a
factorized model as in Eq.~(\ref{eq:factorized}) on many random
subsets of stocks of variable size $N$. One would
naively expect that the typical strength of direct impact propagators ($G^{ij}$ with
$i=j$) be roughly constant regardless of $N$. On the other hand, cross propagators
($i\not = j$) are expected to decrease as $N^{-1}$ or faster, in order to avoid that their contribution ends up dominating
over direct impact when $N\rightarrow\infty$.

The results are shown in Fig.~\ref{scaling}. We can see that indeed
$\left\langle G^{ij}\right\rangle_{i = j}$ decreases only very slightly with $N$ (fitting it to $k_1 N^{-\nu_1}$ yields $k_1=0.36$ and $\nu_1=0.04$), while we
get an excellent fit of cross terms $\left\langle G^{ij}\right\rangle_{i\not = j} = {k_2}\big({1+\frac{N}{N_2}}\big)^{-1}$, with $k_2=0.06$
and $N_2 = 24$ (see dashed lines in Fig.~\ref{scaling}). This suggests a total asymptotic contribution of the off-diagonal propagators equal to $k_2 N_2 \approx 1.5$, to be compared to an average diagonal contribution of $\approx 0.3$.

We have also made a fit on the average of off-diagonal elements,
conditioned such that $i$ and $j$ are in the same sector s: $\left\langle G^{ij}\right\rangle_{i\not = j; \mathrm{s}(i) = \mathrm{s}(j)} = {k_3}\big({1+\frac{N}{N_3}}\big)^{-\nu_3}$ gives $k_3 = 0.078$, $N_3 = 10.4$ and $\nu_3 = 0.54$.
Pairwise cross-impact is naturally stronger in this case, than between two
randomly selected stocks, since they are more likely to have a high
correlation. Nevertheless, understanding how $\nu_3$ should behave is
more delicate, as it requires estimating how the sizes of the sectors themselves
scale with $N$.

%%%%%%%%%%%%%%%%%%%%%%%%%

\section{Estimators of $G$: structure and statistical significance}
\label{sec:dimensional_reduction}

\subsection{The models}
\label{sec:models}

The model defined in Eq.~(\ref{def_G}) is a very general
object, that is described by a propagator $\Prop^{ij}_\tau$ of dimension
$N^2 \times T$, plus a covariance matrix $\cret^{ij}_W$ of
dimension $N^2$.
Such an abundance of parameters results in the impossibility to
estimate reliably the individual entries of $\Prop^{ij}_\tau$ with the
data in our possession. Only the aggregated statistics of
$\Prop^{ij}_\tau$ have been found statistically significant, see
Fig.~\ref{G_matrix} and Table~\ref{score_tab} below.
We have thus decided to use a fully non-parametric estimation only in
order to extract the main qualitative features of data. In order to estimate
the interaction strengths $\Prop^{ij}$ and investigate their structure in
a more robust fashion, we have considered lower dimensional models.

More precisely, we have used three models in order to fit the
propagators and we have compared their performance:
\begin{description}
\item[Fully non-parametric] The most general propagator model is
  specified the $N^2 (T+1)$ parameters defining
  Eq.~(\ref{def_G}). This corresponds to the absence of any prior about
  the structure of the  $\Prop^{ij}_\tau$.
\item[Factorized] A simpler model is obtained under the assumption
$ \Prop^{ij}_\tau = G^{ij}\phi_{\tau} $,
where $\phi_{\tau}$ given by Eq.~(\ref{time_dep}). The dimensionality
of the model is then reduced to $2N^2 + T$.
\item[Homogeneous] The simplest, non-trivial linear model for cross-impact
  is obtained by assuming
  \begin{eqnarray}
    \Prop^{ij} &=& \delta^{ij}\Prop^{\mathrm {diag}}
                   +(1-\delta^{ij})\Prop^{\mathrm{off}},  \\
    \cret^{ij}_W &=& \delta^{ij}\cret^{\mathrm {diag}}_W
                       +(1-\delta^{ij})\cret^{\mathrm{off}}_W \ ,
  \end{eqnarray}
  so as to capture a single collective mode of the return
  covariance, corresponding to global market moves. The dimensionality of the
  model in this case is $4+T$. The estimators for
this model are reported in the Appendix, and yield $G^{\mathrm{diag}} = 0.29$ and $G^{\mathrm{off}}=0.0046$, consistent with the average diagonal and off-diagonal values of the previous model (see \ref{sec:factorized-model}).
\end{description}
All these models can be calibrated by minimizing their negative
log-likelihood under a Gaussian assumption for the residuals $w^i_t$:
\begin{equation}
  \label{eq:logl} - \ln \mathcal{L} = \frac{T}{2} \ln \det \cret_{W}
+ \frac 1 2 \sum_{i,j,t} w^i_t w^j_t \left(\cret_W^{-1}\right)^{ij},
\end{equation} allowing  us to compute estimators for both the propagators
$\Prop^{ij}_\tau$ and the residual covariance matrix
$\cret^{ij}_W$.\footnote{Note that the Gaussian assumption can be
relaxed, as the Generalized Method of Moments employed for example in Refs.~\cite{bouchaud2004fluctuations,bouchaud2006random,eisler2012price}
yields the same estimators that we have derived.  Nevertheless, we
choose the Gaussian assumption for the residuals in order to have
closed-form results for the residuals and the likelihood function.}
In this way, the estimated covariance matrix of the residual $\hat
\cret^{(W)}$ itself can be used in order to build metrics for the
quality of the fit, and check how well the results generalize
out-of-sample.

\subsection{Discussion}
\label{sec:discussion}
The effort of fitting the propagator model under the different models
described above can be justified by two different perspectives. On
the one hand from the \emph{statistical} point of view, it is desirable to avoid
overfitting, so to have a robust model that generalizes
well out-of-sample. This enables us to predict the future
covariation of prices given the imbalances. On the other hand, from
the \emph{informational} point of view, one might prefer to
compress the structure of the interaction in a small number of
informative parameters, rather than dealing with a larger set of
more anonymous coefficients.
In order to quantitatively address these
points, we have chosen to inspect the behavior of the residuals and of
the likelihoods in all the three models, by defining three types of
scores.
The first two scores assess how well one is able to describe the
fluctuations along the diagonal and the off-diagonal parts of the
return covariance matrix (thus, they specify particular axes of the
matrix $\cret^{(W)}$ in which we are interested:
\begin{eqnarray}
  \label{eq:3} \mathcal{R}^{\mathrm{diag}} &=& \frac{\sum_{i} \hat
\cret_W^{ii}}{\sum_{i}\cret^{ii}_0} = \frac 1 N \sum_{i} \hat
\cret_W^{ii} , \\ 
\mathcal{R}^{\mathrm{off}} &=& \frac{\sum_{i\neq j}
|\hat \cret^{ij}_W|}{\sum_{i\neq j}|\cret^{ij}_0|}.
\end{eqnarray}
Alternatively, the likelihood function automatically considers the
fluctuations along the eigenmodes of $\cret^{(W)}$, as its value is
uniquely fixed by the eigenvalues of the residual covariance:
\begin{eqnarray}
\mathcal{R}^{\ln \mathcal{L}} &=& - \frac{\ln \mathcal L}{NT} =
\frac{1}{2} \left( 1 - \frac 1 N \log \det \hat \cret_W \right).
\end{eqnarray}
Table~\ref{score_tab} compares these scores for an in-sample period (2012) and
an out-of-sample one (2013), in order to assess how each of the model
generalizes to yet unseen data. {Note that the in-sample scores are consistent with the results of Fig.~\ref{explained_cov_fig} at lag $\tau=1$, indicating that the metrics that we have chosen provides a very conservative estimate of the model performance due to the increase of the predictive power with lag.} We find that:
\begin{itemize}
\item All the in-sample scores improve by increasing the
  complexity of the models, as expected due to the fact that the
  models are nested. On the contrary, the good
  in-sample performance of the fully non-parametric model does not
  generalize out-of-sample. The scores displayed
  by the lower dimensional models are roughly the same in and
  out-of-sample, thus validating the practical use of the
  factorized and homogeneous propagator models.
\item The quality in the reconstruction of the covariance of returns,
  measured by $\mathcal R^{\mathrm{off}}$, is better than the one of the
  variance. While the factorized model explains around 20\% of the
  variance, it accounts for more than 50\% of the covariance. This
  is compatible with the findings of Ref. \cite{hasbrouck2001common}
  using a model with purely permanent impact.
\item While both the factorized model and the homogeneous one have
  good out-of-sample performance, it is interesting to notice that thes
  factorized model has a consistently better $\mathcal R^{\mathrm{off}}$
  score. This is consistent with our previous results (see
  Fig.~\ref{overlap}), indicating that the directional structure of
  the matrix $\Prop^{ij}$ is statistically significant, and allows one to
  explain a consistent part of the return covariation.
\end{itemize}
The good out-of-sample performance also indicates that heterogeneities
in the temporal behaviour of the propagator discussed in
Section~\ref{sec:mult-prop-model} are weak enough for these models to
generalize well across years. 

\begin{table}
\begin{center}
\caption{\label{score_tab}Table of scores for the three models
  described in Sec.~\ref{sec:dimensional_reduction}.}
\footnotesize
\begin{tabular}{@{}l|lll|lll}
\br
& \multicolumn{3}{c|}{in-sample (2012)} &
                                         \multicolumn{3}{c}{out-of-sample
                                         (2013)} \\[.5em]
\ \ \ \ \ \ Model&   $\mathcal R^{\mathrm{diag}}$  & $\mathcal R^{\mathrm{off}}$ & $\mathcal{R}^{\ln \mathcal{L}}$ &            $\mathcal R^{\mathrm{diag}}$  & $\mathcal R^{\mathrm{off}}$ & $\mathcal{R}^{\ln \mathcal{L}}$  \\
\mr
Non-parametric &     0.437 &   0.185 & \   0.466 &                 2.08 &   1.312 & \  1.187 \\
Factorised     &     0.748 &   0.374 &  \ 0.744 &                 0.79 &   0.454 &  \ 0.762 \\
Homogeneous    &     0.819 &   0.484 & \  0.841 &                0.786 &   0.628 &  \  0.81 \\
\br
\end{tabular}
\end{center}
\end{table}
\normalsize

\section{Conclusions}
\label{sec:conclusions}

The treatment of cross-impact in the existing literature has been
scarce at best \cite{hasbrouck2001common, pasquariello2013strategic,
  wang2016average, wang2016cross}, despite its importance -- in our
opinion -- to correctly estimate the liquidation costs of a
diversified portfolio. In this work we have attempted to give a more
complete picture of such effects by decomposing them using a simple,
linear propagator approach. Our dynamical model explains rather well
the off-diagonal elements of the correlation matrix, which makes us
conclude that to a large extent, {\it cross-correlations between different stocks are mediated by trades}. 
Market makers/HFT liquidity providers learn from correlated
order flow on multiple instruments, and adjust their prices at a
portfolio level. This allows them to better adapt to global movements
in the market, and to reduce the amount of adverse selection they are faced with. Such
an observation is underpinned by the good fit of our homogeneous
model, where each stock reacts to the total, net order flow of the
others. This focus of market makers/HFT on their net inventory is
consistent with the idea that being uniformly long or short across
stocks is much more risky than to be long-short by same gross amount
in a diversified fashion. \smallskip

In the present study we took an empirical, descriptive point of view
regarding price reaction and market maker behavior. In particular, we
have  disregarded the strong implications that these results have in the
context of optimal execution, that will be the object of a forthcoming
paper~\cite{mastromatteo2016liquidity}.\smallskip

We wish to thank R. Benichou, J. Bun, A. Darmon, L. Duchayne, S. Hardiman, J.~Kockelkoren, J. de Lataillade, C.-A. Lehalle, F. Patzelt,  E. S\'eri\'e and B. T\'oth  for very fruitful discussions.

\clearpage

\bibliographystyle{iopart-num}
\bibliography{bibs}

\clearpage

\appendix

\section{Models}
\label{app:models}

\begin{figure}[t!]
\begin{center}
\includegraphics[width=1\columnwidth]{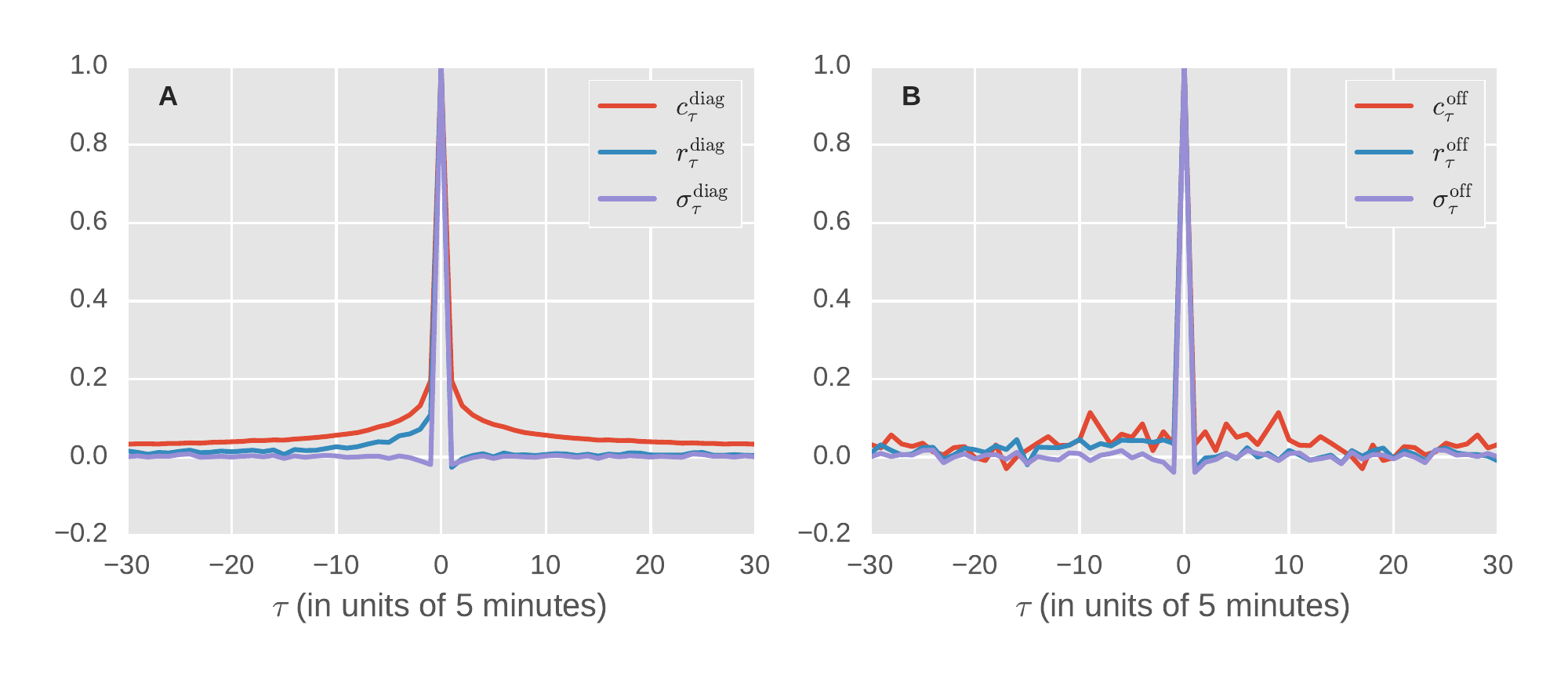} \caption{Plot of the diagonal (a) and off-diagonal (b) means of the lagged sign correlation function, the returns lagged correlation function and the differential response function.} \label{smallcaps}
\end{center}
\end{figure}

It is important to mention that while the results in this paper are presented
with integrated response functions $\resp^{ij}_\tau$ and
propagators $\Prop^{ij}_{\tau}$, all propagator inversions have
been done with differential response functions $r^{ij}_\tau$. This is
consistent with the idea that such quantities have a decaying
asymptotic behaviour in contrast with their integrated counterparts
and thus suffer less from the cut-off effect
\cite{serie2010unpublished, eisler2011models}. The integrated
propagator was then computed by using the relation $\Prop^{ij}_{\tau}
= \sum_{\tau'=1}^{\tau} g^{ij}_{\tau'}$. Figure~\ref{smallcaps} displays the
diagonal and off-diagonal means of the lagged sign correlation
function, the lagged return correlation function and the differential response function.

\subsection{Fully non-parametric model}
\label{sec:fully-non-parametric}
The maximization of the likelihood function (\ref{eq:logl}) of the model in the fully non-parametric case yields a
well-known matrix equation for the propagator:
{
\begin{equation}
  \label{eq:nonparam_g}
  \hat r^{ij}_{\tau} = \sum_{k} \sum_{\tau'=0}^{T-1} \, \hat \prop^{ik}_{\tau'} \, 
 \hat \csgn^{kj}_{\tau,\tau'} \ ,
\end{equation}
that is defined in terms of the (biased) estimators for, respectively, the differential response and the sign correlation:
\begin{eqnarray}
  \label{}
  \hat r^{ij}_\tau &=& \frac 1T \sum_{t,t'=1}^T x^i_t\varepsilon^j_{t'}\delta(t-t'-\tau) \\
   \hat c^{ij}_{\tau,\tau'}  &= & \frac 1T \sum_{t,t',t''=1}^T \varepsilon^i_{t'}\varepsilon^j_{t''}\delta(t-t'-\tau)\delta(t-t''-\tau') \ . 
\end{eqnarray} 
In order to reduce noise and facilitate matrix inversion, we've assume stationarity, so that we define the following stationary estimator for the sign correlation, in which we enforce a Toeplitz structure:
\begin{eqnarray}
  \label{}
   \hat c^{ij}_{\tau - \tau'}  &= & \frac 1T \sum_{t,t'=1}^T \varepsilon^i_{t}\varepsilon^j_{t'}\delta(t-t'-\tau) \ .
\end{eqnarray} 
so that Eq.~(\ref{eq:nonparam_g}) becomes a simpler convolution.
}
The estimator of the residuals is also straightforward to compute:
\begin{equation}
  \label{eq:nonparam_s}
  \hat \cret^{ij}_W = \frac{1}{T} \sum_{t=1}^T w^{i}_t w^j_t \, .
\end{equation}
The total number of parameters to estimate under this method is $N^2
(T+1)$, while the computational bottleneck results from the inversion
of the block-Toeplitz matrix $\csgn^{ij}_\tau$ appearing in Eq.~(\ref{eq:nonparam_s}).

\pagebreak

\subsection{Factorized model}
\label{sec:factorized-model}
The assumption of a propagator of the form
\begin{eqnarray}
\Prop^{ij}_\tau &=& G^{ij}\phi_{\tau} \, ,
\end{eqnarray}
where $\phi_{\tau}$ is given by Eq.~(\ref{time_dep}), results in a
simpler estimation of $N^2 +T$ parameters for the kernel and $N^2$
parameters for the residuals. The estimator for the propagator is
found by solving:
\begin{equation}
  \label{eq:factor_g}
  \hat G^{ij} = A\left(B^T\right)^{-1} \, , 
\end{equation}
where one has preliminarily defined:
\begin{eqnarray}
  \label{eq:7}
  A^{ij} &=&  \sum_\tau \hat R^{ij}_\tau \phi_{\tau}  \\
  B^{ij} &=&  \sum_{\tau,\tau'}  \hat c^{ij}_{\tau - \tau'}  \phi_{\tau} \phi_{\tau'} \, ,
\end{eqnarray}
whereas the estimator of the variance is given by the earlier expression~(\ref{eq:nonparam_s}).

\subsection{The homogeneous model}
\label{sec:homogeneous-model}

The estimator for the propagator reads
\begin{eqnarray}
  \label{}
\hat G^{\mathrm{diag}} &=&\frac1N\left(  \frac{A^{\mathrm{M}}}{B^{\mathrm{M}}}+(N-1) \frac{A^{\mathrm{M}}-A^{\mathrm{I}}}{B^{\mathrm{M}}-B^{\mathrm{I}}} \right)\, , \\
\hat G^{\mathrm{off}}  &=& \frac1N \left(\frac{A^{\mathrm{M}}}{B^{\mathrm{M}}}  - \frac{A^{\mathrm{M}}-A^{\mathrm{I}}}{B^{\mathrm{M}}-B^{\mathrm{I}}}\right) \, , 
\end{eqnarray} 
where one has preliminarily defined the market (M) and idiosyncratic (I) means:
\begin{eqnarray}
  \label{}
  A^{\mathrm{M}} &=&  \avg[A^{ij}]     =   \frac1{N^2}\sum_{ij} A^{ij} \, ,    \\
  A^{\mathrm{I}}   &=& \avg[ A^{ii}]   = \frac{\mathrm{Tr}[A]}N  \, ,
\end{eqnarray}
and equivalently for $B^{\mathrm{M}}$ and $B^{\mathrm{I}}$.   The estimator of the variance is given by (we give the inverse of the estimator for simplicity):
\begin{eqnarray}
  \label{}
 \left( \hat \cret_{W}\right)^{-1,\mathrm{diag}} &=& \frac1N\left( \lambda^{\mathrm M}+(N-1)\lambda^{\mathrm I}  \right) \, ,\\
 \left( \hat \cret_{W}\right)^{-1,\mathrm{off}}     &=&   \frac1N\left( \lambda^{\mathrm M}-\lambda^{\mathrm I}  \right) \, .
\end{eqnarray}
We have also introduced
   \begin{eqnarray}
  \label{}
\lambda^{\mathrm M}&=&\frac1N \left[ \avg[\sigma_0]  - \frac{{A^{\mathrm M}}^2}{B^{\mathrm M}}    \right]^{-1} \, , \\
\lambda^{\mathrm I}  &=& \frac{N-1}{N}\left[  1- \avg[\sigma_0] +\frac{\left(  A^{\mathrm M}-A^{\mathrm I} \right)^2}{ B^{\mathrm M}-B^{\mathrm I} }  \right]^{-1} \ .
\end{eqnarray}

\end{document}